\documentclass[pra,aps,showpacs,preprint]{revtex4}
\usepackage{amssymb}
\usepackage{amsmath}

\usepackage{graphicx}

\newcommand*{\erfc}{\mathop{\mathrm{erfc}}}
\newcommand*{\erf}{\mathop{\mathrm{erf}}}
\newcommand*{\re}{\mathop{\mathrm{Re}}}

\begin{document}

\title{Point process model of 1/f noise versus a sum of Lorentzians}
\author{B. Kaulakys}
\author{V. Gontis}
\author{M. Alaburda}
\affiliation{Institute of Theoretical Physics and Astronomy, Vilnius University, \\
A.~Go\v stauto 12, LT-01108 Vilnius, Lithuania}

\begin{abstract}
To be published in \emph{Phys. Rev. E} (2005).
\vspace{1cm}

\noindent We present a simple point process model of $1/f^{\beta}$ noise, covering 
different values of the exponent $\beta $. The signal of the model
consists of pulses or events. The interpulse, interevent, interarrival, 
recurrence or waiting times of the signal are described by the general Langevin 
equation with the multiplicative noise and stochastically 
diffuse in some interval resulting in the power-law distribution. Our model 
is free from the requirement of a wide distribution of relaxation times and 
from the power-law forms of the pulses. It contains only one 
relaxation rate and yields $1/f^{\beta }$ spectra in a wide range of frequency. 
We obtain explicit expressions for the power spectra and present numerical 
illustrations of the model. 
Further we analyze the relation of the point process model of $1/f$ noise with the Bernamont-Surdin-McWhorter model, representing the signals as a sum of the uncorrelated components. We show that the  point process model is complementary to the 
model based on the sum of signals with a wide-range distribution of the 
relaxation times. In contrast to the Gaussian distribution of the signal intensity of 
the sum of the uncorrelated components, the point process exhibits asymptotically 
a power-law distribution of the signal intensity. 
The developed multiplicative point process model of $1/f^{\beta}$ noise may be used for modeling and analysis of stochastic processes in different systems with the power-law distribution of the intensity of pulsing signals. 
\end{abstract}

\maketitle

\section{Introduction}

\noindent $1/f$ fluctuations are widely found in nature, i.e., the power 
spectra $S(f)$ of a large variety of physical, biological,
geophysical, traffic, financial and other systems at low frequencies $f$
have $1/f^{\beta }$ (with $0.5\lesssim \beta \lesssim 1.5$) behavior 
\cite{Press78,rev1,Thu97,Scien95}. Widespread occurrence of signals
exhibiting such a behavior suggests that a generic mathematical explanation 
of $1/f$ noise might exist. The generic origins of two 
popular noises: white noise (no correlation in time, $S(f)\sim 1/f^0$) and 
Brownian noise (no correlation between increments, $S(f)\sim 1/f^2$) are 
very well known and understood. It should be noted, that the Brownian motion is the integral of white noise and that operation of integration of the signal increases the 
exponent by 2 while the inverse operation of differentiation decreases it by 
2. Therefore, $1/f$ noise can not be obtained by simple procedure of 
integration or differentiation of such convenient signals. 
Moreover, there are no simple, even linear stochastic differential equations 
generating signals with $1/f$ noise. Recently we derive a stochastic nonlinear 
differential equation for the signal exhibiting 
$1/f$ noise in any desirably wide range of frequency \cite{KR04}. 
The physical interpretation of this highly nonlinear equation is not so clear and straightforward as that of the linear Langevin equation, generating the Brownian 
motion of the signal with $1/f^2$ spectrum. 
Therefore, $1/f$ noise is often represented as 
a sum of independent Lorentzian spectra with a wide range of relaxation times 
\cite{b1}. Summation or integration of the Lorentzians with the appropriate 
weights may yield $1/f$ noise. 

Not long ago a simple analytically solvable model of $1/f$ noise has been 
proposed \cite{Ka98}, analyzed \cite{KM98,Ka2000}, and generalized \cite{KM99}. 
The signal in the model consists of pulses or series of events (a point 
process). The interpulse times of the signal stochastically diffuse about some 
average value. The process may be described by an autoregressive iteration 
with a very small relaxation. The proposed model reveals one of the possible 
origins of $1/f$ noise, i.e., random increments of 
the time interval between the pulses (the Brownian motion in the time axis), 
sometimes resulting in the clustering of the signal pulses \cite{Ka98,KM98,KM99}. 

The power spectral density of such point process may be expressed as 
\begin{equation}
S(f)\simeq 2\bar{I}^{2}\bar{\tau}P_{k}\left( 0\right) /f.  \label{S(f)}
\end{equation}
Here $\bar{\tau}=\left\langle \tau _{k}\right\rangle$ is the expectation of 
the interpulse time $\tau _{k}=t_{k+1}-$ $t_{k}$, with $\left\{
t_{k}\right\}$ being the sequence of the pulses occurrence times or arrival times $t_{k}$, 
whereas $P_{k}\left( \tau _{k}\right)$ is a steady state distribution 
density of the interpulse time $\tau _{k}$ in $k$-space and $\bar{I}$ is 
the average intensity of the signal 
\begin{equation}
I\left( t\right) =\sum_{k}A_{k}\left( t-t_{k}\right) .  \label{I(t)}
\end{equation}
Function $A_{k}\left( t-t_{k}\right) $ represents the shape of the $k$-pulse 
of the signal in the region of the pulse occurrence time $t_{k}$. 

It is easy to show that the fluctuations and shapes of $A_{k}\left( t-t_{k}
\right)$ for sharp pulses mainly influence the high frequency power 
spectral density. Therefore, in a low frequency region we can restrict our 
analysis to the noise originated from the correlations between the occurrence 
times $t_{k}$. Then we can simplify the signal to the point process 
\begin{equation}
I\left( t\right) =\bar{a}\sum_{k}\delta \left( t-t_{k}\right)   \label{Ip(f)}
\end{equation}
with $\bar{a}$ being an average contribution to the signal of one pulse or 
one particle when it crosses the section of observation. 

Point processes arise in different fields, such as physics, economics, cosmology, ecology, neurology, seismology, traffic flow, signaling and telecom networks, audio streams, and 
Internet (see, e.g., \cite{Thu97,Grun,Earth,tr,p1} and references herein). 
The proposed point process model \cite{Ka98,KM98,KM99} can been modified and useful 
for the modeling and analysis of self-organized systems \cite{p2}, atmospheric 
variability  \cite{p3}, large flares from Gamma-ray Repeaters in astronomy 
\cite{p4}, particles moving in viscous fluid \cite{p5}, dynamical percolation 
\cite{p6}, $1/f$ noise observed in cortical neurons and earthquake data \cite{p7}, 
financial markets \cite{GoKa03}, cognitive experiments \cite{Scien95,p8}, 
the Parkinsonian tremors \cite{p9}, and time intervals production in tapping and 
oscillatory motion of the hand \cite{HM2004}. 

The analytically solvable model and its generalizations \cite{Ka98,KM98,Ka2000,KM99} 
contain, however, some shortage of generality, i.e., it results only in exact $1/f$ 
(with $\beta =1$) noise and only if $P_{k}\left( \tau _{k}\right)
\simeq const$ when $\tau _{k} \rightarrow 0$. 
On the other hand, the numerical analysis of the 
generalized model with different restrictions for diffusion of the 
interpulse time $\tau _{k}$ reveals $1/f^{\beta }$ spectra with $1\lesssim
\beta \lesssim 1.5$ \cite{KM99}. 

The aims of this paper are to generalize the analytical model seeking to 
define the variety of time series exhibiting the power spectral density 
$S(f)\sim 1/f^{\beta }$ with $0.5\lesssim \beta \lesssim 2$ and to analyze the 
relation of the point process model with the Bernamont-Surdin-McWhorter 
model \cite{b1}, representing the signal as a sum of the appropriate 
signals with the different rates of the linear relaxation. 

\section{Power spectral density of the point process}

\noindent The point process is primarily and basically defined by the occurrence times $t_{k}$. 
The power spectral density of the point process (\ref{Ip(f)}) may be expressed as 
\cite{Ka98,KM98,KM99}
\begin{eqnarray}
S\left( f\right) &=&\lim \limits_{T\rightarrow \infty }\left\langle \frac
2T\int\limits_{t_i}^{t_f}\int\limits_{t_i}^{t_f}I\left( t^{\prime }\right)
I\left( t^{^{\prime \prime }}\right) e^{i\omega\left( t^{^{\prime \prime
}}-t^{\prime }\right) }{\rm d}t^{\prime }{\rm d}t^{^{\prime \prime
}}\right\rangle  \notag \\
&=&\lim \limits_{T\rightarrow \infty }\left\langle \frac{2\bar{a}^2}
T\left| \sum_{k=k_{\min }}^{k_{\max }}e^{-i\omega t_k}\right| ^2\right\rangle \notag \\
&=&\lim \limits_{T\rightarrow \infty }\left\langle \frac{2\bar{a}^2}T\sum_{k=k_{\min
}}^{k_{\max }}\sum_{q=k_{\min }-k}^{k_{\max }-k}e^{i\omega\Delta \left(
k;q\right) }\right\rangle  \label{r1}
\end{eqnarray} 
where $T=t_f-t_i\gg \omega ^{-1}$ is the observation time,  
$\omega=2\pi f$, and 
\begin{equation}
\Delta \left( k;q\right) \equiv t_{k+q}-t_{k}=\sum_{i=k}^{k+q-1}\tau _{i}
\end{equation}
is the difference between the pulses occurrence times $t_{k+q}$ and $t_{k}$. 
Here $k_{\min }$ and $k_{\max }$ are minimal and maximal values of
index $k$ in the interval of observation $T$ and 
the brackets $\left\langle ...\right\rangle $ denote the averaging over 
realizations of the process. 

It should be stressed that the spectrum is related to the underlying process and not to a realization of the process \cite{r1,r2}. Therefore, the averaging over realizations of the process is essential. Without the averaging over the realizations we obtain the squared modulus of the Fourier transform of the data, i.e., the periodogram which is fluctuating wildly and its variance is almost independent of $T$ \cite{r1,r2}. For calculation of the 
power spectrum of the actual signal without the averaging over the realizations one should use the well-known procedures of the smoothing for spectral estimations \cite{r1,r2,Marpl,Stoica}. 

Equation (\ref{r1}) may be rewritten as 
\begin{equation}
S\left( f\right) =2\bar{a}^2\bar{\nu }+\lim \limits_{T\rightarrow \infty
}\left\langle \frac{4\bar{a}^2}T\sum_{q=1}^N\sum_{k=k_{\min }}^{k_{\max }-q}\cos
\left[ \omega\Delta \left( k;q\right) \right] \right\rangle   \label{r2} 
\end{equation} 
where $N=k_{\max }-k_{\min }$ and 
\begin{equation*}
\bar{\nu }=\frac{1}{\bar{\tau}}=\left\langle \lim \limits_{T\rightarrow \infty } 
\frac{N+1}T\right\rangle  
\end{equation*}
is the mean number of pulses per unit time. The first term in the right-hand-site 
of Eq. (\ref{r2}) represents the shot noise, 
\begin{equation}
S_{{\rm shot}}\left( f\right) =2\bar{a}^2\bar{\nu }=2\bar{a} \bar{I}, \label{r3}
\end{equation}
with $\bar{I} =\bar{a}\bar{\nu}$ 
being the average signal. 

Eqs. (\ref{r1})-(\ref{r3}) may be modified as 
\begin{equation}
S\left( f\right) =2\bar a^2\sum\limits_{q=-N}^N\left( \bar{\nu }-\frac{\left| 
q\right| }T\right) \chi _{\Delta \left( q\right) }\left( \omega\right)  \label{r4} 
\end{equation}
and used for evaluation of the power spectral density of the non-stationary process or for the process of finite duration, as well. Here 
\begin{equation}
\chi _{\Delta \left( q\right) }\left( \omega\right) =\overline{\left\langle e^{i\omega 
\Delta \left( q\right) }\right\rangle }=\int\limits_{-\infty }^{+\infty }e^{i\omega 
\Delta \left( q\right) }\Psi_q \left( \Delta \left( q\right) \right) d\Delta 
\left( q\right)  \label{r5} 
\end{equation}
is the characteristic function of the distribution density $\Psi_q \left( \Delta \left( q\right) \right) $ of $\Delta \left( q\right)$, a definition $\Delta \left( q\right) =-\Delta \left( -q\right) =\Delta \left( k;q\right) $ is introduced, and the brackets $\overline{\left\langle \ldots \right\rangle }$ denote the averaging over realizations of the process and over the time (index $k$) \cite{KM98,KM99}. For the non-stationary process or process of the finite duration one should use the real distribution $\Psi_q \left( \Delta \left( q\right) \right)$ with the finite interval of the variation of $\Delta \left( q\right)$ or calculate the power spectra directly according to Eq. (\ref{r1}). 

When the second sum of Eq. (\ref{r4}) in the limit $T\rightarrow \infty$, due to the decrease of the characteristic function $\chi _{\Delta \left( q\right) }\left( \omega\right)$ for finite $\omega$ and large $q$, approaches to zero, 
\begin{equation*}
\lim \limits_{T\rightarrow \infty }\frac{1}T\sum\limits_{q=-N}^N\left| q\right| \chi _{\Delta \left( q\right) }\left( \omega\right) \rightarrow 0, 
\end{equation*}
we have from Eq. (\ref{r4}) the power spectrum in the form 
\begin{equation}
S\left( f\right) =\lim\limits_{T\rightarrow \infty }\left\langle \frac{2
\bar{a}^{2}}{T}\sum\limits_{k,q}{e}^{i\omega\Delta \left(
k;q\right) }\right\rangle=2\bar{I}^{2}\bar{\tau}\sum\limits_{q=-N}^N\chi _{\Delta \left( q\right) }\left( \omega\right).   \label{Sp(f)}
\end{equation}
 
\section{Stochastic multiplicative point process}

\noindent According to the above analysis, the power spectrum of the point process signal is 
completely described by the set of the interpulse intervals 
$\tau _{k}=t_{k+1}-$ $t_{k}$. Moreover, the low frequency 
noise is defined by the statistical properties of the signal at
large-time-scale, i.e., by the fluctuations of the time difference $\Delta
\left( k;q\right)$ at large $q$, determined by the slow dynamics of the 
average interpulse interval $\tau _{k}\left( q\right) =\Delta \left(
k;q\right) /q$ between the occurrence of pulses $k$ and $k+q$. 
In such a case quite generally the dependence of the interpulse time $\tau _{k}$ on the 
occurrence number $k$ may be described by the general Langevin equation with 
the drift coefficient $d\left( \tau _{k}\right) $ and a multiplicative noise 
$b\left( \tau _{k}\right) \xi \left( k\right) $, 
\begin{equation}
\frac{d\tau _{k}}{dk}=d\left( \tau _{k}\right) +b\left( \tau _{k}\right) \xi
\left( k\right) .  \label{Lang1}
\end{equation}
Here we interpret $k$ as continuous variable while the white Gaussian noise $
\xi \left( k\right) $ satisfies the standard condition 
\begin{equation*}
\left\langle \xi \left( k\right) \xi \left( k^{\prime }\right) \right\rangle
=\delta \left( k-k^{\prime }\right)
\end{equation*}
with the brackets $\left\langle ...\right\rangle $ denoting the averaging 
over the realizations of the process.
Equation (\ref{Lang1}) we understand in Ito interpretation. 

Perturbative solution of Eq. (\ref{Lang1}) in the vicinity of $\tau _{k}$ yields 
\begin{equation}
\tau _{k+j}\simeq \tau _{k}+d\left( \tau _k\right) j+b\left( \tau _k\right) \int\limits_k^{k+j}\xi \left( l\right) dl,  \label{r11}
\end{equation}
\begin{equation}
\Delta \left( k;q\right)=\sum_{i=k}^{k+q-1}\tau _{i}
\simeq \int\limits_0^q\tau _{k+j}dj\simeq \tau _kq+d\left( \tau _k\right) 
\frac{q^2}2+b\left( \tau _k\right) \int\limits_0^qdj\int\limits_k^{k+j}\xi 
\left( l\right) dl.  \label{r12}
\end{equation}
After integration by parts we have 
\begin{equation}
\Delta \left( k;q\right) =\tau _{k}q+d\left( \tau _{k}\right) \frac{q^{2}}{2}
+b\left( \tau _{k}\right) \int\limits_{k}^{k+q}\left( k+q-l\right) \xi \left(
l\right) dl,  \label{Delt}
\end{equation}
\begin{equation}
\left\langle \Delta \left( k;q\right) \right\rangle =\tau _{k}q+d\left( \tau
_{k}\right) \frac{q^{2}}{2}.  \label{DeltV}
\end{equation}

Analogously, in the same approximation we can obtain and the variance $\sigma _{\Delta }^{2}(k;q)=\left\langle \Delta \left( k;q\right)^2 \right\rangle-\left\langle \Delta \left( k;q\right) \right\rangle^2$ of the time difference $\Delta \left( k;q\right)$, 
\begin{equation} 
\sigma _{\Delta }^{2}(k;q)=b^2\left( \tau _{k}\right)\frac{q^{3}}{3}.  \label{VarDelta}
\end{equation}

\subsection{Power spectral density}

\noindent Substituting Eqs. (\ref{Delt}) and (\ref{DeltV}) into Eq. (\ref{Sp(f)}) and
replacing the averaging over $k$ by the averaging over the distribution of the
interpulse times $\tau _{k}$ we have the power spectrum 
\begin{eqnarray}
S\left( f\right)  &=&4\bar{I}^{2}\bar{\tau}\int\limits_{0}^{\infty }d\tau
_{k}P_{k}\left( \tau _{k}\right) \re
\int\limits_{0}^{\infty }dq\exp \left\{ i\omega\left[ \tau _{k}q+d\left( \tau
_{k}\right) \frac{q^{2}}{2}\right] \right\}   \notag \\
&=&2\bar{I}^{2}\frac{\bar{\tau}}{\sqrt{\pi }f}\int\limits_{0}^{\infty }P_{k}
\left( \tau _{k}\right) \re\left[ {e}^{
-i\left( x-\frac{\pi }{4}\right) }\erfc\sqrt{-ix
}\right] \frac{\sqrt{x}}{\tau _{k}}d\tau _{k}  \label{Sp2}
\end{eqnarray}
where $x=\pi f\tau _{k}^{2}/d\left( \tau _{k}\right) $. 

The replacement of the averaging over $k$ and over realizations of the process by the averaging over the distribution of the interpulse times $\tau_{k}$, $P_{k}\left( \tau_{k}\right)$, is possible when the process is ergodic. Ergodicity is usually a common feature of the stationary process described by the general Langevin equation \cite{Gar85}. Therefore, we will consider the stationary processes of diffusion of the interpulse time $\tau_k$ described by Eq. (\ref{Lang1}) and restricted in the finite interval the motion. Such restrictions may be introduced as some additional conditions to the stochastic equation. The similar restrictions, however, may be fulfilled by introducing some additional terms into Eq.~(\ref{Lang1}), corresponding to the diffusion in some ``potential well'', as in paper \cite{KR04}.

Approach (\ref{Sp2}) is the improvement of the simplest model of the pure $1/f$ noise 
\cite{Ka98,KM98} taking into account the second, drift, term 
$d\left( \tau _{k}\right)q^2/2$ in expression for $\Delta \left( k;q\right)$. Note, that 
for $d\left( \tau _{k}\right) \rightarrow 0$ from Eq. (\ref{Sp2}) we recover the known 
result (\ref{S(f)}).

According to Eqs. (\ref{S(f)}), (\ref{r1}) and (\ref{Sp2}) the small interpulse 
times and the clustering 
of the pulses make the greatest contribution to $1/f^{\beta }$ noise. 
The power-law spectral density is very often related with the power-law behavior of other characteristics of the signal, such as autocorrelation function, probability densities and other statistics, and with the fractality of the signals, in general \cite{Thu97,b8,Mand97,Mand99,b11,Heavy03,b10}. 
Therefore, we investigate the power-law 
dependences of the drift coefficient and of the distribution density on the
time $\tau _{k}$ in some interval of the small interpulse times, i.e., 
\begin{equation}
d\left( \tau _{k}\right) =\gamma \tau _{k}^{\delta },\quad P_{k}
\left( \tau _{k}\right) =C\tau _{k}^{\alpha },\quad \tau _{\min }\leq
\tau _{k}\leq \tau _{\max }  \label{P}
\end{equation}
where the coefficient $\gamma $ represents the rate of the signal's nonlinear relaxation
and $C$ has to be defined from the normalization. 

The power-law distribution of the interpulse, interevent, interarrival, recurrence or 
waiting time is observable in different systems from physics, astronomy and seismology to 
the Internet, financial markets and neural spikes (see, e.g., \cite{Thu97,p1,p2,r3} 
and references herein). 

One of the most direct applications of the model described by Eq. (\ref{P}), perhaps, is for the modeling of the computer network traffic \cite{p1} with the spreading of the packets of the requested files in the Internet traffic and exhibiting the power-law distribution of the inter-packet time. The modeling of these processes is under way. 

Because of the divergence of the power-law distribution and requirement of the stationarity of the process the stochastic diffusion may be realized over a certain range of the variable    $\tau _{k}$ only. Therefore, we restrict the diffusion of $\tau_k$ in the interval $[\tau_{\min},\tau_{\max}]$ with the appropriate boundary conditions. Then the steady state solution of the stationary Fokker-Planck equation with a zero flow corresponding to Eq. (\ref{Lang1}) is \cite{Gar85} 
\begin{equation}
P_{k}\left( \tau _{k}\right) =\frac{C}{b^{2}\left( \tau _{k}\right)}\exp
\left\{ 2\int\limits_{\tau _{\min }}^{\tau _{k}}\frac{d\left( \tau \right) }
{b^{2}\left( \tau \right)}d\tau \right\}.  \label{Stead}
\end{equation}
For the particular power-law coefficients $d\left(\tau_k\right)$ and $b\left(\tau_k\right)$ (see, e.g., Eq.(\ref{Tau})) we can obtain the power-law stationary distribution density (\ref{P}).

Then equations (\ref{Sp2}) and (\ref{P}) yield the power spectra with different slopes 
$\beta $, i.e., 
\begin{equation}
S\left( f\right) =\frac{2\bar{I}^{2}}{\sqrt{\pi }\left( 2-\delta \right) f}
\left( \frac{f_{0}}{f}\right) ^{\frac{\alpha }{2-\delta }}I_{\kappa }\left(
x_{\min },x_{\max }\right) ,  \label{Sp3}
\end{equation}
\begin{equation}
I_{\kappa }\left( x_{\min },x_{\max }\right) =\re
\int\limits_{x_{\min }}^{x_{\max }}{e}^{-i\left( x-\frac{\pi }{4}
\right) }\erfc\left( \sqrt{-ix}\right) x^{\kappa
}dx.  \label{I}
\end{equation}
Here $\kappa =\frac{\alpha }{2-\delta }-
\frac{1}{2}$, $x_{\min }=f/f_{2}$, $x_{\max }=f/f_{1}$, 
\begin{equation}
f_{0}=\frac{\gamma }{\pi }\left( C\bar{\tau}\right) ^{\frac{2-\delta }{
\alpha }},\quad f_{1}=\frac{\gamma }{\pi \tau _{\max }^{2-\delta }},\quad
f_{2}=\frac{\gamma }{\pi \tau _{\min }^{2-\delta }}.  \label{f0}
\end{equation}
Note that $f_{0}$ is indefinite when $\alpha \rightarrow 0$, however, 
$f_{0}^{\frac{\alpha }{2-\delta }}$ is definite and converges to 
$C\bar{\tau}$ in this limit.

We note the special cases of the power spectral density (\ref{Sp3}).

(i) $f_{1}\ll f\ll f_{2}$, $-1<\kappa <1/2$, 
\begin{equation}
S\left( f\right) =\frac{\Gamma \left( 1+\kappa \right) \bar{I}^{2}}{\sqrt{
\pi }\left( 2-\delta \right) \cos \left[ \left( \kappa /2+1/4\right) \pi 
\right] f}\left( \frac{f_{0}}{f}\right) ^{\kappa +\frac{1}{2}},  \label{Sben}
\end{equation}
i.e., $S\left( f\right) \sim 1/f^{1+\frac{\alpha }{2-\delta }}$ and $
S\left( f\right) \sim 1/f$ for $\alpha =0$, in accordance with Eq. (\ref{S(f)}).

(ii) $f\ll f_{1}$, $\kappa >-1$, 
\begin{equation}
S\left( f\right) =\frac{\bar{I}^{2}}{\left( 1+\alpha -\delta /2\right) }
\left( \frac{f_{0}}{f_{1}}\right) ^{\frac{\alpha }{2-\delta }}\sqrt{\frac{2}{
\pi f_{1}f}},  \label{Smaz}
\end{equation}
i.e., for very low frequencies $S\left( f\right) \sim 1/\sqrt{f}$.

(iii) $f\gg f_{2}$, $\kappa <1/2$, 
\begin{equation}
S\left( f\right) =\frac{\bar{I}^{2}}{\sqrt{\pi }\left( 2-\alpha -\delta
\right) }\left( \frac{f_{0}}{f_{2}}\right) ^{\frac{\alpha }{2-\delta }}\frac{
f_{2}}{f^{2}},  \label{Sdid}
\end{equation}
i.e., for high frequencies $S\left( f\right) \sim 1/f^{2}$. 

For very high frequencies $f\gg \tau _{\max }^{-1}$, however, we can not replace 
the summation in Eq. (\ref{Sp(f)}) by the integration. Then from Eqs. (\ref{r2}) 
or (\ref{Sp(f)}) one gets the shot noise $S\left( f\right)=2\bar{a}\bar{I}$, 
Eq. (\ref{r3}). 

Equations (\ref{Sp3}) and (\ref{Sben})-(\ref{Sdid}) reveal that the
proposed model of the stochastic multiplicative point process may result in 
the power-law spectra over several decades of low frequencies with the slope 
$\beta $ between $0.5$ and $2$.

The simplest and well-known process generating the power-law probability 
distribution function for $\tau _{k}$ is a multiplicative stochastic process 
with $b\left( \tau _{k}\right)=\sigma \tau _{k}^{\mu}$ and $\delta =2\mu -1$, 
written as  \cite{Go02}
\begin{equation}
\tau _{k+1}=\tau _{k}+\gamma \tau _{k}^{2\mu -1}+\sigma \tau _{k}^{\mu
}\varepsilon _{k}.  \label{Tau}
\end{equation}
Here $\gamma $ represents the relaxation of the signal, while $\tau _{k}$
fluctuates due to the perturbation by normally distributed uncorrelated 
random variables $\varepsilon _{k}$ with a zero expectation and unit variance 
and $\sigma $ is a standard deviation of the white noise. According to 
Eq. (\ref{Stead}) the steady state solution of the stationary Fokker-Planck equation 
with a zero flow, corresponding to Eq. (\ref{Tau}), gives the power-law probability 
density function for $\tau _{k}$ in the $k$-space 
\begin{equation}
P_{k}\left( \tau _{k}\right)=\frac{1+\alpha }{\tau _{\max }^{1+\alpha }-\tau _{\min }^{1+\alpha }}\tau_{k}^{\alpha}, \quad  
\alpha =\frac{2\gamma }{\sigma ^2}-2\mu.  \label{r21}
\end{equation}
The power spectrum for the intermediate $f$, $f_1\ll f\ll f_2$, according to 
Eq. (\ref{Sben}) is 
\begin{equation}
S(f)=\frac{\left( 2+\alpha \right) \left( \beta -1\right) \bar a^2\Gamma \left( \beta -1/2\right) }{\sqrt{\pi }\alpha \left( \tau _{\max }^{2+\alpha }-\tau _{\min }^{2+\alpha }\right) \sin \left( \pi \ \beta /2\right) }\left( \frac \gamma \pi \right) 
^{\beta -1} \frac 1{f^\beta }   \label{r22}
\end{equation}
where 
\begin{equation} 
\beta =1+\frac \alpha {3-2\mu }, \quad \frac{1}{2}<\beta<2.   \label{r30}
\end{equation}
For $\mu =1$ we have a completely multiplicative point process when 
the stochastic change of the interpulse time is proportional to itself. 
Multiplicativity is an essential feature of the financial time series, economics, 
some natural and physical processes \cite{Sa00}.

Another case of interest is with $\mu =1/2$, when the Langevin equation in the 
actual time takes the form 
\begin{equation}
\frac{d\tau }{dt}=\gamma \frac{1}{\tau }+\sigma \xi \left( t\right),
\label{Taudt}
\end{equation}
i.e., the Brownian motion of the interpulse time with the linear relaxation of 
the signal $I\simeq \bar{a}/\tau $.

\begin{figure}[p]
\begin{center}
\includegraphics[width=0.7\textwidth]{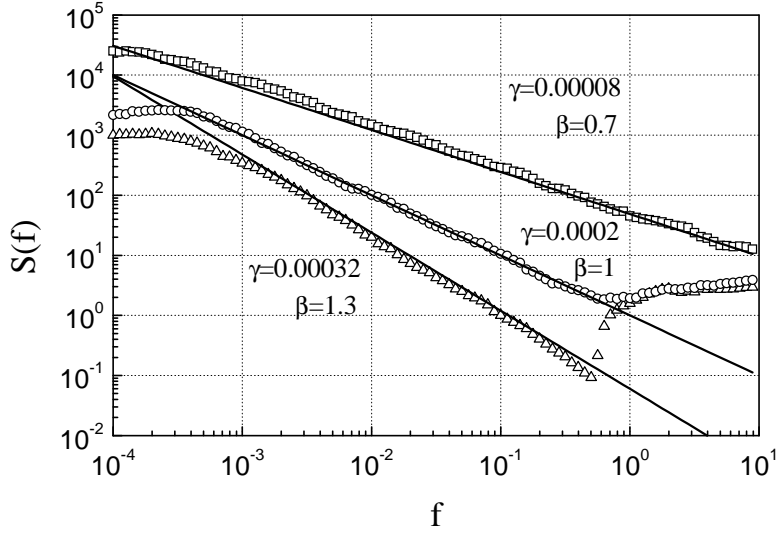}
\end{center}
\vspace{-25pt}
\caption{Power spectral density (\ref{r1}) vs frequency of the signal generated 
by Eqs. (\ref{Ip(f)}) and (\ref{Tau}). Numerical simulations are averaged over 10 realizations 
of $N=10^6$ pulse sequences with the parameters $\protect\bar{a}=1$, 
$\protect\mu =1/2$, $\protect\sigma =0.02$ and different relaxations of the 
signal $\protect\gamma $. We restrict the diffusion of the interpulse time 
in the interval $\protect\tau _{\min }=10^{-6}$, $\protect\tau _{\max }=1$ 
with the reflective boundary condition at $\protect\tau _{\min } $ and 
transition to the white noise, $\protect\tau _{k+1}=\protect\tau _{\max }+
\protect\sigma \protect\varepsilon _{k}$, for $\protect\tau _{k}>\protect\tau
_{\max }$. The straight lines represent the analytical results according to 
Eq. (\ref{r22}).}
\label{fig:1}
\end{figure}

\begin{figure}[p]
\begin{center}
\includegraphics[width=0.7\textwidth]{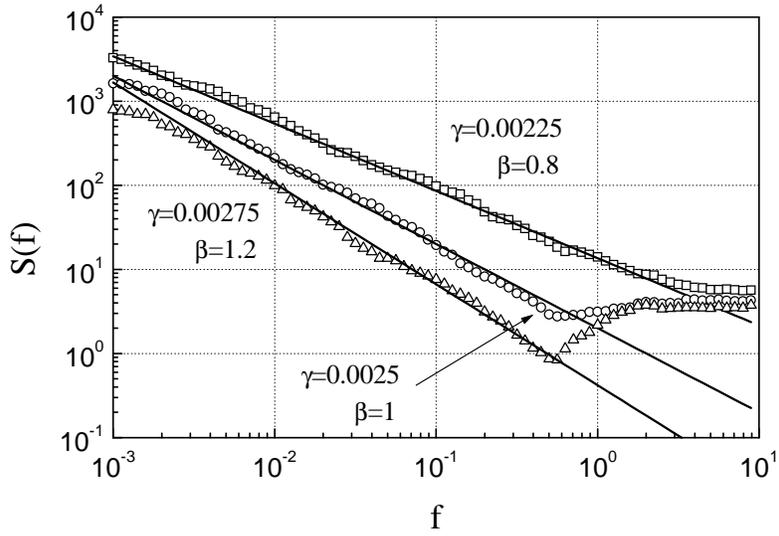}
\end{center}
\vspace{-25pt}
\caption{The same as in Fig. \ref{fig:1} but for $\protect\mu =1$, $\protect\sigma 
=0.05$ and different parameters $\protect\gamma $.}
\label{fig:2}
\end{figure}

Figures \ref{fig:1} and \ref{fig:2} represent the spectral densities (\ref{r1}) 
with the different slopes $\beta $ of the signals generated numerically according 
to Eqs. (\ref{Ip(f)}) and (\ref{Tau}) for the different parameters of the model. 
We see that the simple iterative equation (\ref{Tau}) with the 
multiplicative noise produces the signals with the power spectral density
of different slopes, depending on the parameters of the model. The agreement of the 
numerical results with the approximate theory is quite good. 

It should be noted that the low frequency noise is insensitive to the small 
additional fluctuations of the particular occurrence times $t_{k}$. 
Therefore, we can interpret that Eqs. (\ref{Lang1}), (\ref{Tau}) and (\ref{Taudt}) 
describe the slow diffusive motion of the average interpulse time, superimposed by some additional randomness.

On the other hand, the numerical investigations have shown that the proposed model is
stable with respect to variation of dynamics of the interpulse time $\tau _{k}$. The substitution of the reflecting boundaries for $\tau _{k}$ by an appropriate confining potential as in Ref. \cite{KR04} does not change the result. 

\subsection{Distribution density of the signal intensity}

\noindent The origin for appearance of $1/f$ fluctuations in the point process model
described by Eqs. (\ref{I(t)})-(\ref{Taudt}) is related with the slow, Brownian 
fluctuations of the interpulse time $\tau _{k}$ as a function of the 
pulse number $k$, when the average interpulse time $\tau _{k}(q)$ is a
slowly fluctuating function of the arguments $k$ and $q$. In such a case transition 
from the occurrence number $k$ to the actual time $t$ according to 
the relation $dt=\tau_{k}dk$ yields the probability distribution density 
$P_{t}(\tau_{k})$ of $\tau_{k}$ in the actual time $t$, 
\begin{equation}
P_{t}(\tau_{k})=P_{k}(\tau_{k})\tau_{k}/\bar{\tau}. \label{p5}
\end{equation}

The signal averaged over the time interval $\tau_{k}$ according to Eq. (\ref{Ip(f)}) 
is $I=\bar{a}/\tau_{k}$. Therefore, the distribution density of the intensity of the 
point process (\ref{Ip(f)}) averaged over the time interval $\tau_{k}$ is 
\begin{equation}
P(I)=\frac{\bar{a}\bar{I}}{I^{3}}P_{k}\left(\frac{\bar{a}}{I}\right). \label{p6}
\end{equation}

If $P_{k}(\tau_{k})\simeq const$ when $\tau_{k}\to 0$ (the condition for the exhibition for
the pure $1/f$ noise in the point process model) the distribution density 
of the signal is
\begin{equation}
P(I)\sim I^{-3}. \label{p7}
\end{equation}

For the generalized multiplicative processes (\ref{Ip(f)}), (\ref{Lang1}), and (\ref{P}) 
we have from Eqs. (\ref{r21}) and (\ref{p6}) the distribution density of the signal 
intensity  
\begin{equation}
P(I)=\frac{\lambda -1}{\tau _{\max }^{\lambda -1}-\tau _{\min }^{\lambda -1}}\frac{\bar a^{\lambda -1}}{I^\lambda },\;\lambda=3+\alpha. \label{p7a}
\end{equation}

The power-law distribution of the signals is observable in a large variety of systems 
ranging from earthquakes to the financial time series \cite{Thu97,Earth,GoKa03,b8,Mand97,Mand99,b11,Heavy03,b10,Go02,Gop98}.

One of the simplest models generating the Brownian fluctuations of the interpulse 
time $\tau_{k}$ is an autoregressive model \cite{Ka98,KM98,KM99} with random increments 
and linear relaxation of the interpulse time, i.e., the model described by the
iterative equation
\begin{equation}
\tau_{k+1}=\tau_{k}-\gamma(\tau_{k}-\bar{\tau})+\sigma\varepsilon_{k}. \label{p8}
\end{equation}
Here $\bar{\tau}$ is the average interpulse time, $\gamma$ is the rate of the linear relaxation, $\{\varepsilon_{k}\}$ denotes the sequence of uncorrelated normally 
distributed random variables with zero expectation and a unit variance and $\sigma$ 
is the standard deviation of this white noise. 
The model (\ref{Ip(f)}), (\ref{Sp(f)}), and (\ref{p8}) then results in the power 
spectral density \cite{KM98}
\begin{equation}
S(f)=\bar{I}^{2}\frac{\alpha_{H}}{f},\; 
\alpha_{H}=\frac{2}{\sqrt{\pi}}Ke^{-K^{2}},\;
K=\frac{\bar{\tau}\sqrt{\gamma}}{\sigma}. \label{p9}
\end{equation}
The distribution density of the intensity of the signal according to Eqs. (\ref{Stead}) and (\ref{p6}) then is 
\begin{equation}
P(I)=\frac{K\bar I^2}{\sqrt{\pi }I^3}\exp \left\{-\frac{\gamma\bar{a}^2}{\sigma^2}\left(\frac{1}{\bar{I}}
-\frac{1}{I}\right)^2\right\}. \label{p7b}
\end{equation}

Restricting the diffusion of the interpulse time $\tau_k$ by the reflective boundary 
condition at $\tau_{min}>0$ and for $\tau_{min}\to 0$ we have the truncated distribution density of the signal intensity 
\begin{equation}
P_r(I)=\frac{2K\bar I^2}{\sqrt{\pi }\left[ 1+\erf\left( K\right) \right] }\exp \left\{ -K^2\left( 1-\frac{\bar I}I\right) ^2\right\} \frac {1}{I^3}, \; I>0. \label{p71} 
\end{equation}

In the asymptotic $I\gg\bar{I}$ and $I\gg 2 K^2\bar{I}$ from Eq. (\ref{p71}) we have 
\begin{equation}
P_r(I)\simeq \alpha_{H}^r\frac{\bar{I}^2}{I^3}\sim \frac{1}{I^3},  \label{p7c}
\end{equation}
i.e., the power-law distribution density of the signal. Here 
\begin{equation}
\alpha_{H}^r=\frac{\alpha_{H}}{1+\erf\left(K\right)}.  \label{p7d}
\end{equation}

The restriction of motion of $\tau _k$ by the reflective boundary condition at $\tau _k=0$ reduces the effective (average) value of  $P_k\left(0\right)=\frac{1}{2}\left[P_k\left(\tau_k\rightarrow +0\right) +P_k\left(\tau_k\rightarrow -0\right) \right]$ in Eq. ({\ref{S(f)}}) and, consequently, the power spectral density approximately 2 times in comparison with the theoretical result ({\ref{p9}}) obtained without the restriction, because $P_k\left(\tau_k\rightarrow -0\right) =0$ for the restricted motion. 
More exactly, in such a case the power spectral density may be expressed by Eq. (\ref{p9}) with $\alpha_{H}^r$ instead of $\alpha_{H}$, i.e., 
\begin{equation}
S_r(f)=\bar{I}^{2}\frac{\alpha_{H}^r}{f}.  \label{p7e}
\end{equation}
\subsection{Correlation function of the point process}

\noindent Correlation function $C(s)$ of the point process (\ref{Ip(f)}) may be 
expressed as 
\begin{equation}
C(s)=\left\langle \frac{\bar{a}^{2}}{T}\sum_{k,q}\delta
(t_{k+q}-t_{k}-s)\right\rangle =\bar{I}\bar{a}\sum_{q}\int\limits_{-\infty }^{+\infty }
\Psi_q \left( \Delta \left( q\right) \right) \delta (\Delta
(q)-s) d\Delta \left( q\right)=\bar{I}\bar{a}\sum_{q}\Psi_q \left( s \right)  \label{p10}
\end{equation}
where the brackets $\langle \ldots \rangle $ denote the averaging over the 
realizations of the process and over time (index $k$) as well. Such 
averaging coincides with the averaging over the distribution of the time 
difference $\Delta (q)$, $\Psi_q \left( \Delta \left( q\right) \right) $. 

From Eq. (\ref{p10}) for the approximation 
\begin{equation}
\Delta (k;q)\equiv t_{k+q}-t_{k}=\sum_{l=k+1}^{k+q}\tau _{l}\simeq \tau
(q)q,\;q\geq 0  \label{p3}
\end{equation}
we have the expression for the correlation function in the simplest 
approximation \cite{KM99} 
\begin{equation}
C(s)\simeq \bar{I}\bar{a}\sum_{q}\int\limits_{\tau_{\min}}^{\tau_{\max}} 
P_{k}(\tau _{k})\delta (\tau _{k}q-s)d\tau_{k}=\bar{I}\bar{a}\delta(s)
+\bar{I}\bar{a}\sum_{q\not=0}P_{k}\left( \frac{s}{q}
\right) \frac{1}{|q|}.  \label{p11}
\end{equation}
Replacing the summation in Eq. (\ref{p11}) by the integration we have the approximate expression for the correlation function of the point processes (\ref{Ip(f)}) and 
(\ref{Lang1}) or (\ref{p8})  
\begin{equation}
C(s)\simeq \bar{I}\bar{a}\int\limits_{0}^{\infty }P_{k}\left( \frac{s}{q}\right) \frac{
dq}{q},\;s\geq 0, \; C(-s)=C(s).  \label{p12}
\end{equation}

\section{Signal as a sum of uncorrelated components}

\noindent As it was already mentioned above, $1/f$ noise is often modeled as 
the sum of the Lorentzian spectra with the appropriate weights of a 
wide range distribution of the relaxation times $\tau^{rel}$. 
It should be noted that the summation of the spectra is allowed only if
the processes with different relaxation times are isolated one from another 
\cite{b1,b5}. For the construction of the signal $I(t)$ with $1/f$ noise
spectrum from the stochastic equations with a wide range distribution of the 
relaxation times (and rates $\gamma _{l}=1/\tau ^{rel}_{l}$) one should
express the signal as a sum of $N$ uncorrelated components \cite{Ka2000} 
\begin{equation}
I(t)=\sum_{l=1}^{N}I_{l}(t)  \label{l1}
\end{equation}
where every component $I_{l}$ satisfies the stochastic differential equation 
\begin{equation}
\dot{I}_{l}=-\gamma _{l}(I_{l}-\bar{I}_{l})+\sigma _{l}\xi _{l}(t).
\label{l2}
\end{equation}
Here $\bar{I}_{l}$ is the average value of the signal component $I_{l}$ , 
$\xi _{l}(t)$ is the $\delta $-correlated white noise, $\langle \xi
_{l}(t)\xi _{l^{\prime }}(t^{\prime })\rangle =\delta _{l,l^{\prime }}\delta
(t-t^{\prime })$, and $\sigma _{l}$ is the intensity (standard deviation) of
the white noise. 

The distribution density $P(I_{l})$ of the component $I_{l}$ is Gaussian 
\begin{equation}
P(I_{l})=\sqrt{\frac{\gamma_{l}}{\pi}}\frac{1}{\sigma_{l}}
\exp \left\{ -\frac{\gamma _l}{\sigma _l^2}\left( I_l-\bar I_l\right) ^2\right\}.
\label{l3}
\end{equation}

The distribution density $P(I)$ of the signal $I(t)$, Eq. (\ref{l1}), expressed as 
a sum of uncorrelated Gaussian components, is Gaussian as well,
\begin{equation}
P(I)=\frac{1}{\sqrt{2\pi}\sigma}
\exp \left\{ -\frac{\left( I-\bar I\right) ^2}{2\sigma ^2}\right\},  \label{l4}
\end{equation}
with the average value $\bar{I}$ and the variance $\sigma^{2}$ expressed as 
\begin{equation}
\bar{I}=\sum_{l}\bar{I}_{l},\; \sigma^{2}=\sum_{l}\frac{\sigma_{l}^{2}}{2\gamma_{l}}.
\label{l5}
\end{equation}

Therefore, the Bernamont-Surdin-McWhorter model based on the sum of signals with a wide 
range distribution of the relaxation times always results in the Gaussian distribution 
of the signal intensity. However, not all signals exhibiting $1/f$ noise are Gaussian 
\cite{rev1}. Some of them are non-Gaussian, exhibiting power-law distribution or even 
fractal \cite{Thu97,b8,Mand97,Mand99,b11,Heavy03,b10}. 
 
Eqs. (\ref{l1}) and (\ref{l2}) result in the
expression for the correlation function of the signal (\ref{l1}), 
\begin{equation}
C(s)=\sum_{l}\frac{\sigma _{l}^{2}}{2\gamma _{l}}e^{-\gamma _{l}s},\;s\geq 0.
\label{l6}
\end{equation}
The correlation function (\ref{l6}) yields the power spectrum 
\begin{equation}
S(f)=\sum_{l}\frac{2\sigma_{l}^2}{\gamma_{l}^{2}+\omega^{2}},\; \omega=2\pi f.
\label{l7}
\end{equation}

Introducing the distribution of the relaxation rates, $g(\gamma )$, we can
replace the summation in Eqs. (\ref{l1}) and (\ref{l5})-(\ref{l7}) by the integration and express 
the power spectrum of the signal (\ref{l1}) as 
\begin{equation}
S(f)=\int\limits_{\gamma _{\min}}^{\gamma _{\max}}\frac{2\sigma ^{2}(\gamma )g(\gamma )}{
\gamma ^{2}+\omega ^{2}}d\gamma =\frac{1}{\pi f}\int\limits_{y_{\min}}^{y_{\max}}\frac{
\sigma ^{2}(\omega y)g(\omega y)}{1+y^{2}}dy.  \label{l8}
\end{equation}
Here $\gamma _{\min}$ and $\gamma _{\max}$ are minimal and maximal values of the relaxation rate, respectively. 

\subsection{Signals with the pure $1/f$ power spectrum} 

\noindent Eq. (\ref{l8}) yields the pure $1/f$ power spectrum only in the case when 
$\sigma ^{2}(\omega y)g(\omega y)=A=const$. In such a case the correlation 
function (\ref{l6}) may be expressed as 
\begin{equation}
C(s)=\frac{A}{2}\int\limits_{\gamma _{\min}}^{\gamma _{\max}}e^{-\gamma s}\frac{d\gamma }{\gamma 
}=\frac{A}{2}\int\limits_{\tau _{\min}^{rel}}^{\tau _{\max}^{rel}}e^{-s/\tau^{rel}}
\frac{d\tau ^{rel}}{\tau^{rel}}  \label{l9}
\end{equation}
while the power spectrum (\ref{l8}) yields 
\begin{equation}
S(f)=\frac A{\pi f}\left[ \arctan \left( \frac{\gamma _{\max }}\omega \right) -
\arctan \left( \frac{\gamma _{\min }}\omega \right) \right] \simeq \frac{A}{2f},\; \gamma _{\min}\ll \omega \ll \gamma _{\max}.  \label{l10}
\end{equation}

For the signal expressed not as a sum (\ref{l1}) but as an average of $N$ uncorrelated 
components,
\begin{equation}
I_{a}(t)=\frac{1}{N}\sum_{l=1}^{N}I_{l}(t),  \label{l11}
\end{equation}
all characteristics (\ref{l3})-(\ref{l10}) are similar, except that the average value
$\bar{I_{a}}$ of the averaged signal (\ref{l11}) is $N$ times smaller than that according to 
Eq. (\ref{l5}), while the expressions for the correlation function $C(s)$, Eqs. (\ref{l6}) 
and (\ref{l9}), for the power spectrum $S(f)$, Eqs. (\ref{l7}), (\ref{l8}), and 
(\ref{l10}), and for the variance $\sigma_{a}^{2}$, Eq. (\ref{l5}), should be divided 
by $N^{2}$, i.e.,
\begin{equation}
\bar{I_{a}}=\frac{1}{N}\sum_{l}\bar{I}_{l}, \;  \sigma_{a}^{2}=\frac{1}{N^2}\sum_{l}\frac{\sigma_{l}^{2}}{2\gamma_{l}},  \label{l11a}
\end{equation}
\begin{equation}
S_{a}(f)\simeq \frac{A}{2N^{2}f},  \label{l12}
\end{equation}
\begin{equation}
C_{a}(s)=\frac 1{2N^2}\int\limits_{\gamma _{\min}}^{\gamma _{\max}}\frac{e^{-\gamma s}}{\gamma }{\sigma^2 \left( \gamma \right)}g(\gamma )d\gamma.  \label{Cg}
\end{equation}

When replacing the summation in Eqs. (\ref{l1}), (\ref{l5})-(\ref{l8}) and (\ref{l11})-(\ref{Cg}) 
by the integration, the distribution density 
of the relaxation rates, $g\left( \gamma \right) $, should be normalized to the number of uncorrelated components $N$, 
\begin{equation}
\int\limits_{\gamma _{\min}}^{\gamma _{\max}}g(\gamma )d\gamma =N.  \label{Nor}
\end{equation}
\noindent 

We see the similarity of expressions (\ref{p12}) and (\ref{Cg}) for the correlation function of the point process model and that of the sum of signals with different relaxation rates, respectively. 
In general, however, different distributions $P_{k}(\tau _{k})$ of
the interpulse time $\tau _{k}$ when $P_{k}(0)\not=0$, e.g., 
exponential, Gaussian and continuous distributions, with the slowly 
fluctuating interpulse time $\tau _{k}$ may result in $1/f$ noise. 
Therefore, the point process model is, in some sense, more general than the model 
based on the sum of the Lorentzian spectra.
 
\begin{figure}[htb]
\begin{center}
\vspace{-15pt}
\includegraphics[width=.5\textwidth]{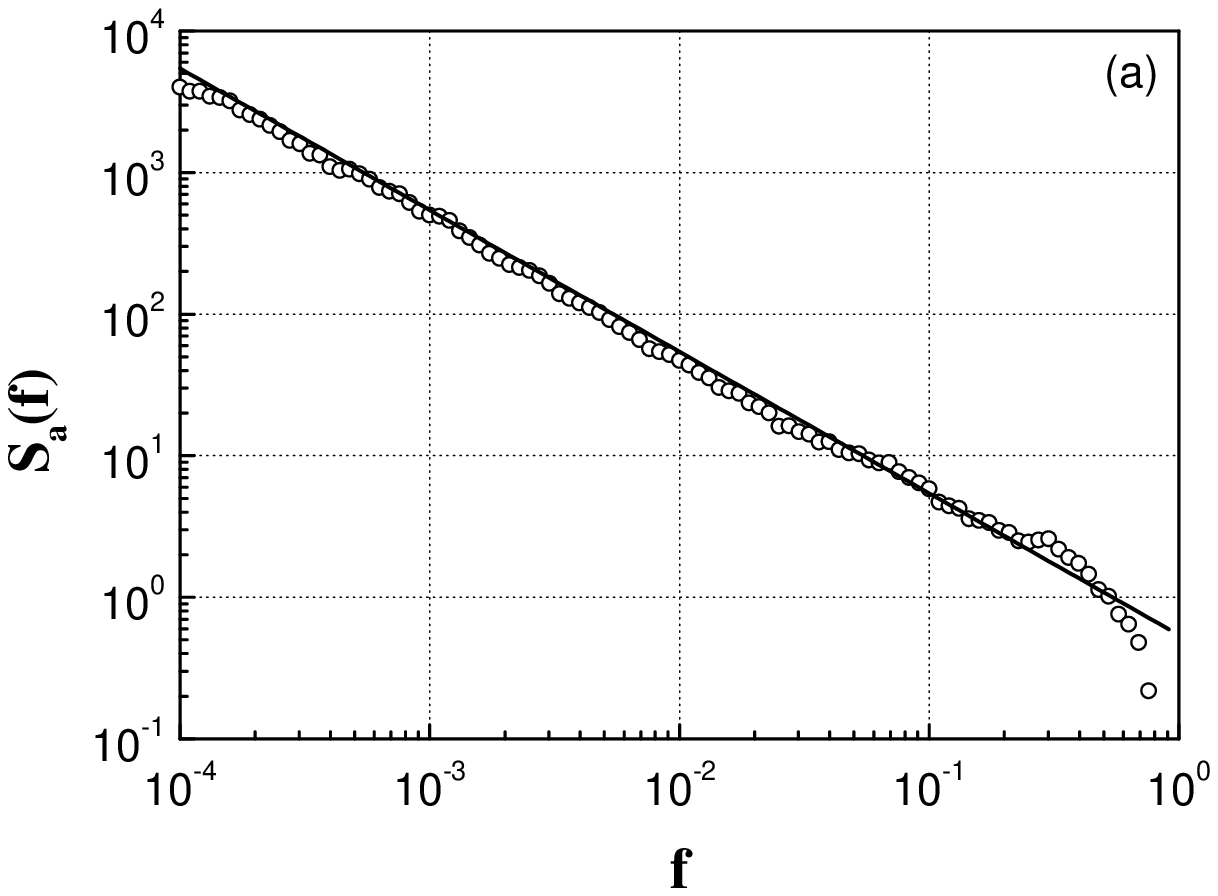}\hspace{-15pt}
\includegraphics[width=.5\textwidth]{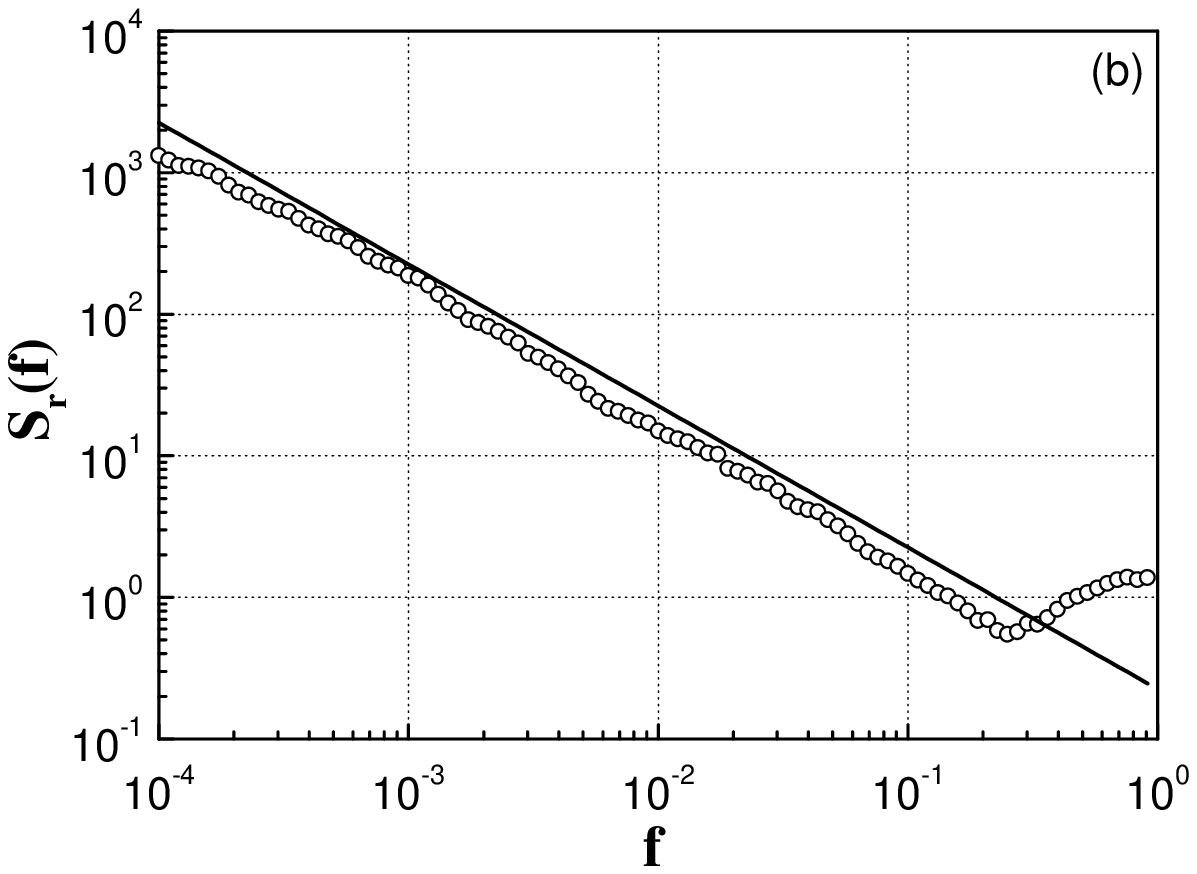}
\includegraphics[width=.5\textwidth]{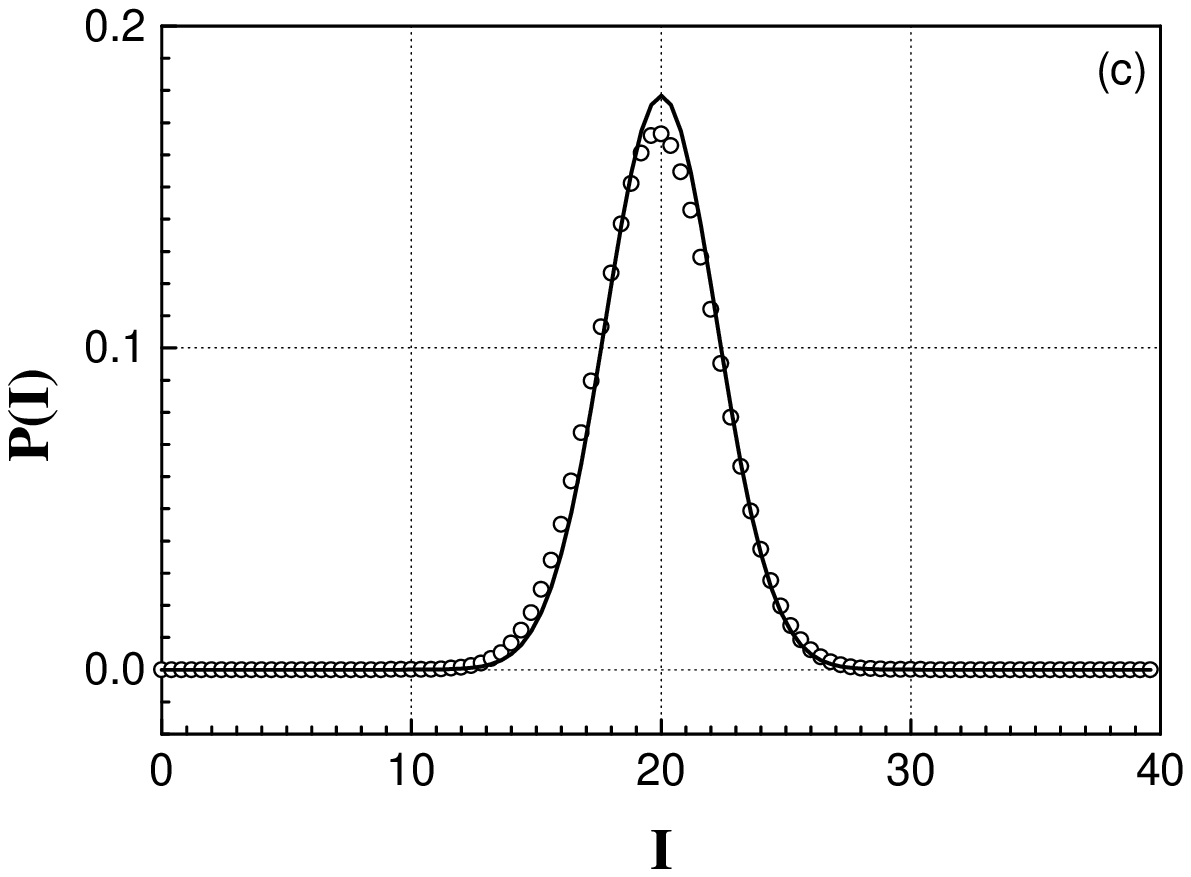}\hspace{-15pt}
\includegraphics[width=.5\textwidth]{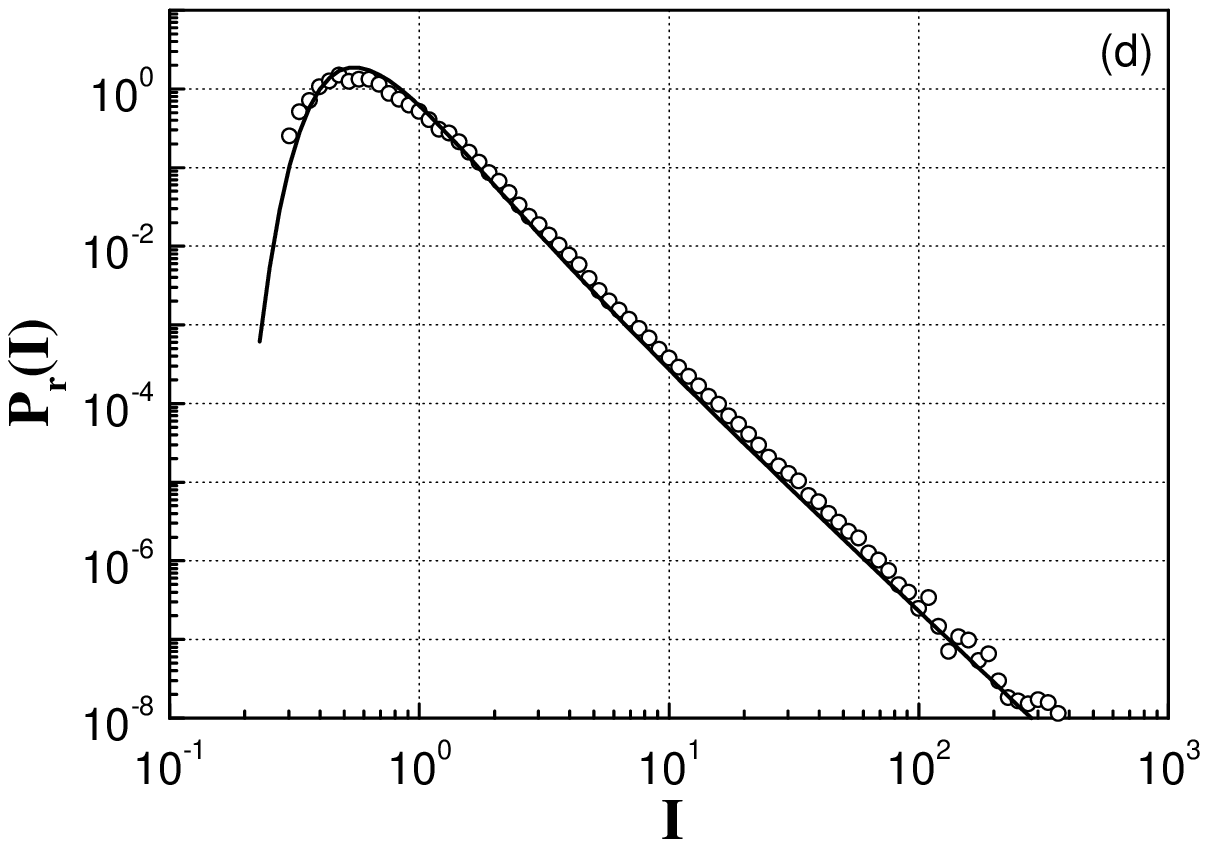}
\end{center}
\par
\vspace{-25pt}
\caption{Power spectra: (a) numerically calculated for the average signal (\ref{l11}) from 
$N=10$ components (\ref{l2}) with $\bar{I}=20$, $\sigma_{l}^{2}(\gamma_{l})g(\gamma_{l})=const$ 
and uniform distribution of $\lg\gamma_{l}$ of $\gamma_{l}$ values in the interval 
$10^{-4}-10^{0}$, i.e., with $g\left( \gamma_l \right) \sim \gamma_l ^{-1}$, $\sigma^2_l 
\left( \gamma_l \right) \sim \gamma_l $, and $\sigma _1\left( \gamma _1\right) =0.1$, 
open circles, in comparison with theoretical results (\ref{l12}), straight line; 
(b) for the point process (\ref{Ip(f)}), (\ref{r1}), and (\ref{p8}) with 
$\bar{a}=1$, $\bar{\tau}=1$, $\sigma=0.01$, and $\gamma=0.0001$ averaged over 10 
realizations of $10^5$ pulse sequences, open circles, in comparison with the theoretical results according to Eq. (\ref{p7e}), straight line. 
(c) and (d) numerically calculated distribution densities of the corresponding signals, open circles, in comparison with the theoretical results (\ref{l4}), (\ref{l11a}), and (\ref{p71}), 
solid lines, respectively.}
\label{spectrum}
\end{figure}

In Figure \ref{spectrum}  the examples of the pure $1/f$ power spectra for the average (\ref{l11}) of signals (\ref{l2}) generated for different relaxation rates $\gamma_l$ 
and with the corresponding intensities of the white noise $\sigma_l^{2}$ and those of the autoregressive point process (\ref{Ip(f)}), (\ref{r1}), and (\ref{p8}) are 
presented together with the distribution densities of the corresponding signals. 
We see the similarity of the spectra but very different distributions 
of the intensity of the signals: the signal of the sum of the Lorentzians 
is Gaussian while that of the point process is approximately of the power-law type, asymptotically $P(I)\sim I^{-3}$. 

\subsection{Signals with the power spectral density of different slopes $\beta$}

\noindent Using the sum of different Lorentzian signals we can generate not only a signal with the pure $1/f$ spectrum but the signal with any predefined slope $\beta$ of $1/f^{\beta}$ power spectral density, as well. Indeed, let us investigate the case when 
\begin{equation}
\sigma^{2}(\gamma)g(\gamma)=A\gamma^{\eta},  
\label{dsp:1}
\end{equation}
where $A$ and $\eta$ are some parameters. 
Substitution of Eq.~\eqref{dsp:1} into Eq.~\eqref{l8} yields the power spectral density 
\begin{multline}
S(f)=\frac{A}{\pi f}\int\limits_{\gamma _{\min}/\omega}^{\gamma _{\max}/\omega}\frac{(\omega y)^{\eta}}{1+y^{2}}dy  \\
=\frac{A}{\omega^{1-\eta}}\left\lbrace
\left[\frac{\gamma _{\max}}{\omega}\right]^{\eta+1}
\Phi\left(-\left[\frac{\gamma _{\max}}{\omega}\right]^{2},1,\frac{\eta+1}{2}\right) \right.\\
\left. -\left[\frac{\gamma _{\min}}{\omega}\right]^{\eta+1}
\Phi\left(-\left[\frac{\gamma _{\min}}{\omega}\right]^{2},1,\frac{\eta+1}{2}\right)\right\rbrace  \label{dsp:2}
\end{multline}
where $\Phi(z,s,a)$ is a Lerch's Phi transcendent. In the limit when $\gamma _{\min}\to 0$ and $\gamma _{\max}\to\infty$ 
we can approximate the power spectral density \eqref{dsp:2} as 
\begin{equation}
S(f)\simeq \frac{\left( 2\pi \right) ^\eta A}{2\cos \left( \pi \eta /2\right) }\frac 1{f^{1-\eta }}, \label{dsp:3}
\end{equation}
i.e., we have the generalization of the result (\ref{l10}). 

For the average signal \eqref{l11} we have 
\begin{equation}
S_{a}(f)\simeq \frac{\left( 2\pi \right) ^\eta A}{2  N^2 \cos \left( \pi \eta /2\right) }\frac 1{f^{1-\eta }}. \label{dsp:4}
\end{equation}

In order to obtain an arbitrary $\beta$ of the $1/f^{\beta}$ power spectral density we should choose in Eq.~\eqref{dsp:1} $\eta=1-\beta$. 

The distribution density $P_{a}(I_{a})$ of the average signal $I_{a}(t)$ is Gaussian 
\begin{equation}
P_{a}(I_{a})=\frac{1}{\sqrt{2\pi}\sigma_{a}}e^{-\frac{(I-\bar{I}_{a})^{2}}{2\sigma^{2}_{a}}} 
\label{dsp:5}
\end{equation}
with the variance $\sigma_{a}^{2}$ expressed as
\begin{equation}
\sigma _a^2=\frac 1{2N^2}\int\limits_{\gamma _{\min }}^{\gamma _{\max }}
\frac{\sigma ^2(\gamma )g(\gamma )}\gamma d\gamma =\frac{A\left( \gamma _{\max }^\eta -
\gamma _{\min }^\eta \right) }{2N^2\eta }.  \label{dsp:6}
\end{equation}

The correlation function in such a case according to Eq. (\ref{Cg}) is 
\begin{equation}
C_a(s)=\frac A{2N^2}\int\limits_{\gamma _{\min }}^{\gamma _{\max }}e^{-\gamma s}
\gamma ^{\eta -1}d\gamma =\frac A{2N^2s^\eta }\left[ \Gamma \left( \eta ,\gamma _
{\min }s\right) -\Gamma \left( \eta ,\gamma _{\max }s\right) \right]   \label{dsp:7}
\end{equation}
where $\Gamma \left( a,z\right) $ is the incomplete gamma function. 
\noindent 
\begin{figure}[htb]
\begin{center}
\vspace{-15pt}
\includegraphics[width=.5\textwidth]{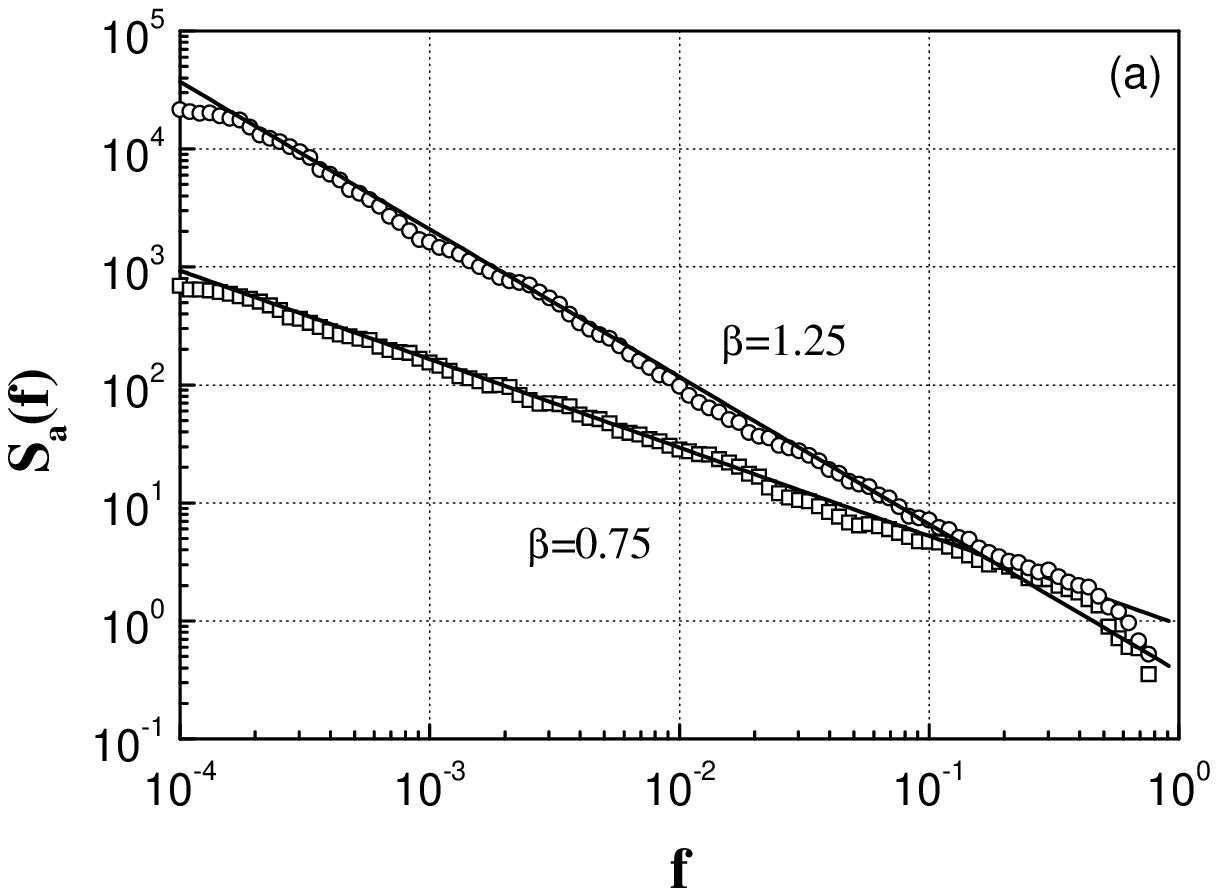}\hspace{-15pt}
\includegraphics[width=.5\textwidth]{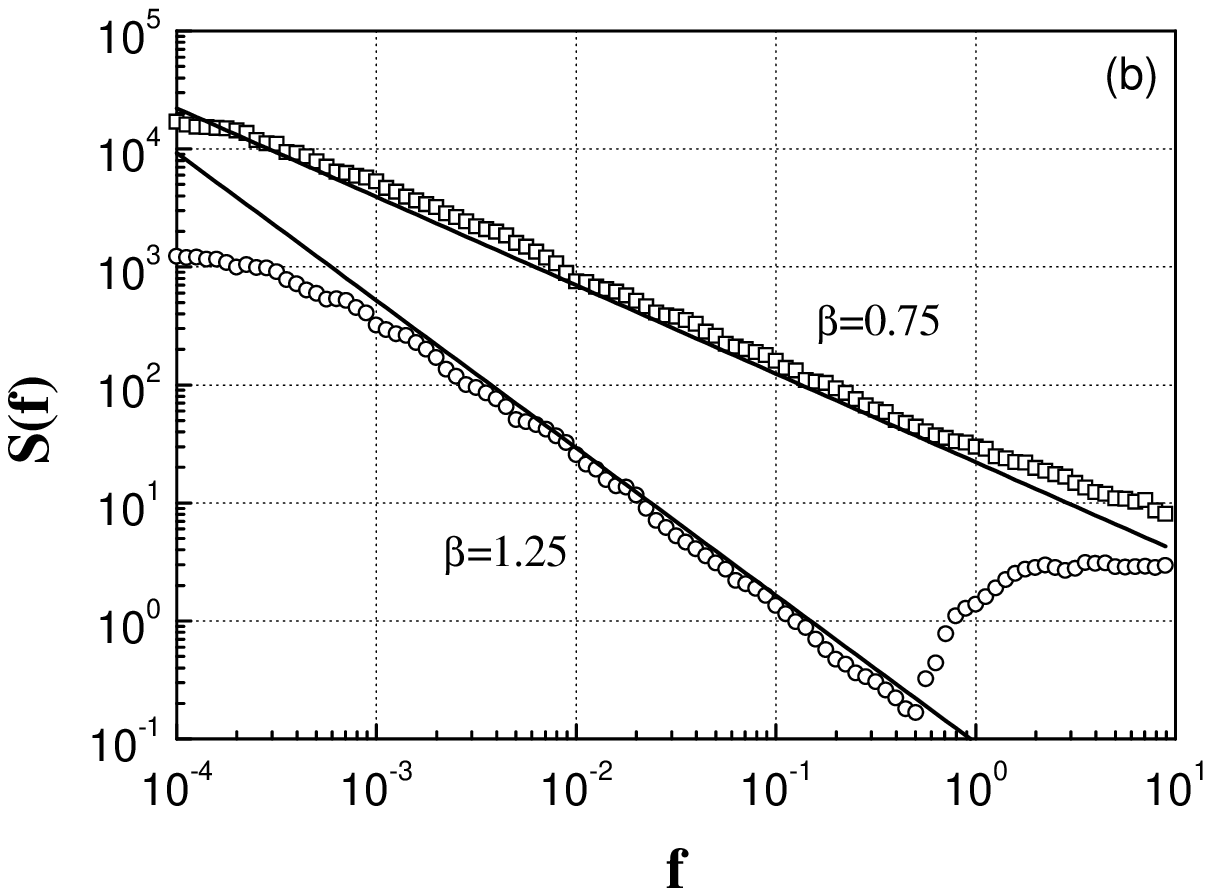}
\includegraphics[width=.5\textwidth]{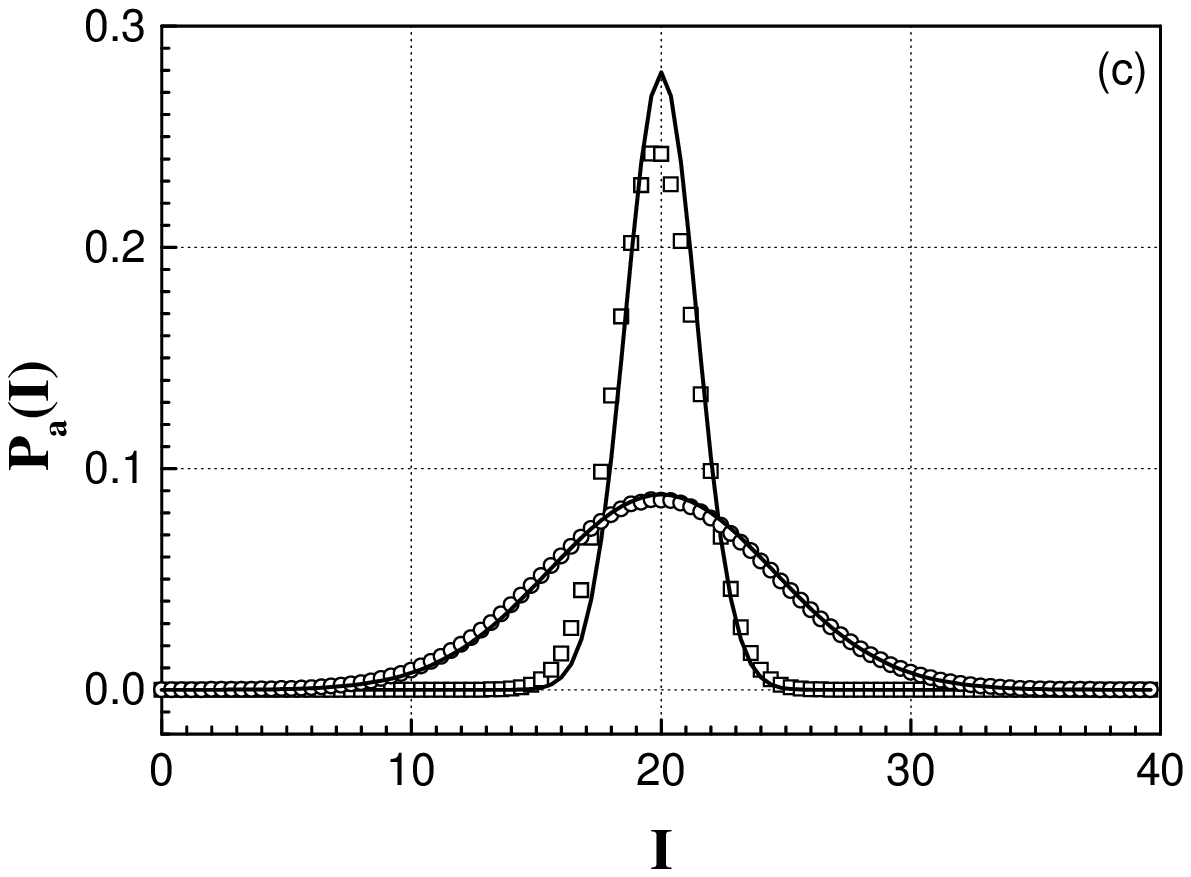}\hspace{-15pt}
\includegraphics[width=.5\textwidth]{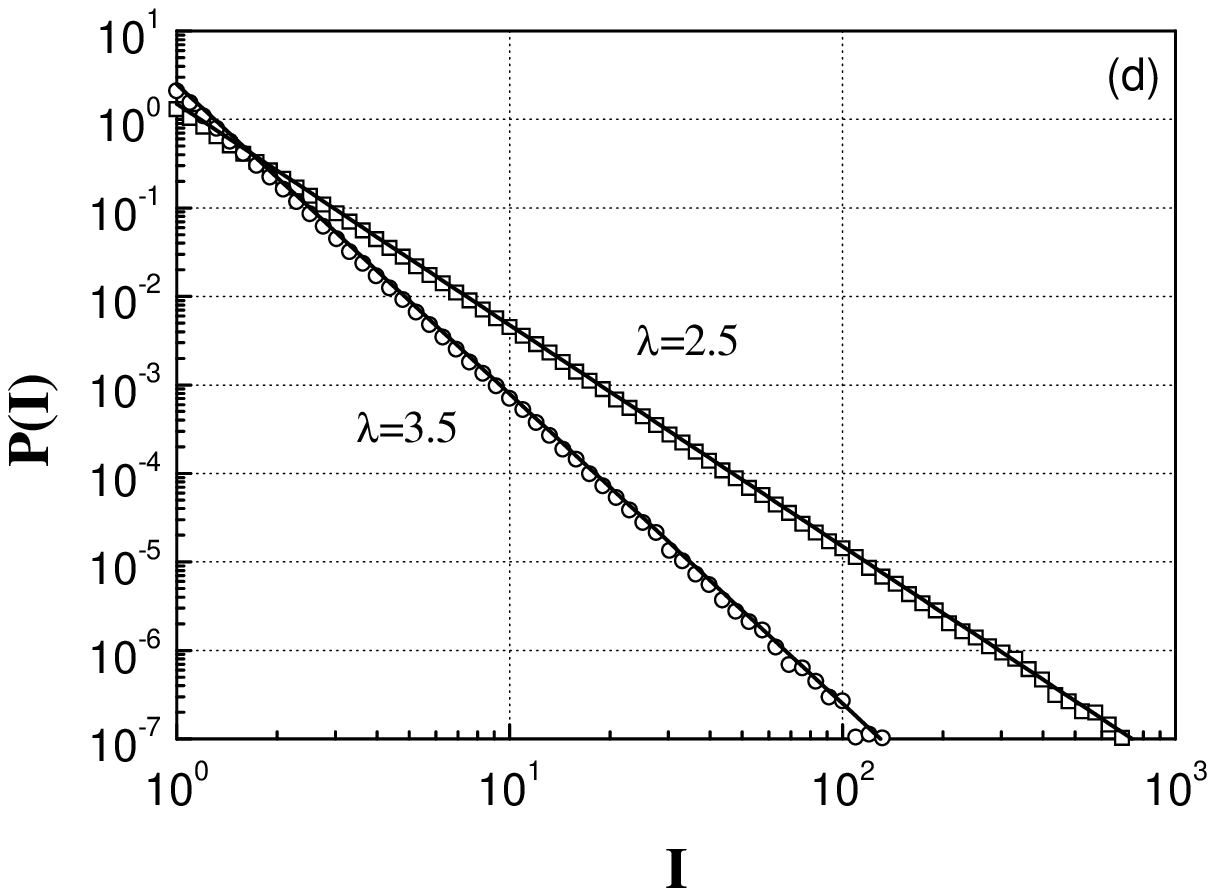}
\end{center}
\par
\vspace{-25pt}
\caption{Power spectra: (a) numerically calculated for signal (\ref{l2}), (\ref{l11}) and (\ref{dsp:1}) from 10 components with $\bar{I}=20$, $A=100$, $\eta=-0.25$, open circles, and $\eta=0.25$, open squares, in comparison with theoretical results (\ref{dsp:4}), straight line; 
(b) for the point process (\ref{Ip(f)}), (\ref{r1}) and (\ref{Tau}) with the parameters $\protect\bar{a}=1$, $\mu=0.5$, $\sigma=0.02$, and $\gamma=0.0001$, open squares, and $\gamma=0.0003$, open circles, averaged over 10 realizations of $10^6$ pulse sequences in comparison with the theoretical results (\ref{r22}), straight lines. 
(c) and (d) numerically calculated distribution densities of the corresponding signals in comparison with the theoretical results (\ref{dsp:5}), (\ref{dsp:6}), and (\ref{p7a}), respectively, solid lines.}
\label{spectrum2}
\end{figure}

Figure \ref{spectrum2} demonstrates the possibility to generate stochastic signals exhibiting similar $1/f^{\beta}$ power spectral densities with different slopes $\beta$ by the summation of signals with different relaxation rates and according to the multiplicative point process model. The distribution densities of the corresponding signals are, however, completely different.

\section{Conclusions}

\noindent The generalized multiplicative point processes (\ref{Ip(f)}), (\ref{Lang1}), (\ref{P}), and (\ref{Tau}) may generate time series exhibiting the power spectral 
density $S(f)\sim1/f^\beta$ with $0.5\lesssim \beta \lesssim 2$, Eqs. (\ref{Sp2}), (\ref{Sben}), and (\ref{r22}), i.e., with the slope observable in a large variety of 
systems. Such spectral density is caused by the stochastic diffusion of the interpulse time, resulting in the power-law distribution. The power-law distribution of the interpulse, interevent, interarrival, recurrence or waiting times is observed in different systems from physics, astronomy and seismology to the Internet, financial markets, neural spikes, and 
human cognition.

Furthermore, the power-law distribution of the interpulse time results in the power-law distribution of the stochastic signal, $P(I)\sim I^{-\lambda}$ with $2\lesssim \lambda \lesssim 4$, i.e., the phenomenon observable in a large variety of processes, from earthquakes to the financial time series, as well. The proposed model relates and connects the power-law autocorrelation and spectral density with the power-law distribution of the signal intensity into the consistent theoretical approach. The generated time series of the model are fractal since they exhibit jointly the power-law probability distribution and the power-law autocorrelation of the signal. 

In addition, we have analyzed the relation of the point process model with the Bernamont-Surdin-McWhorter model of $1/f$ noise, representing the signal as a sum of the appropriate signals with the different rates of the linear relaxation.
From the performed analysis we can conclude that the multiplicative point process model of $1/f$ noise when the signal consisting of pulses with a stochastic motion of the interpulse
time is more general and complementary to the model based on the sum of signals 
with a wide-range distribution of the relaxation times. 
In contrast to the Gaussian distribution of the intensity of sum of the 
uncorrelated components, the point process model generating $1/f$ noise 
exhibits the power-law distribution of the intensity of the signal. 
Moreover, it is free from the requirement of a wide-range distribution of the relaxation times. Obviously, the multiplicative point process model of $1/f^\beta$ noise may be used for modeling and analysis of stochastic processes in different systems exhibiting the pulsing signals.  

\noindent \textbf{Acknowledgments} 

We acknowledge support by the Lithuanian State and Studies Foundation.

\end{document}